\documentclass[a4paper,11pt,dvips]{article}
\usepackage{graphicx}
\usepackage{amsmath}
\usepackage{amssymb}
\usepackage{latexsym}
\usepackage{color}
\usepackage{cite}
\usepackage{makeidx}
\usepackage{float}
\usepackage{multirow}
\usepackage[dvipdfm,colorlinks]{hyperref}
\newcommand{\bra}{\begin{array}}
\newcommand{\era}{\end{array}}
\newcommand{\beq}{\begin{equation}}
\newcommand{\eeq}{\end{equation}}
\newcommand{\bqr}{\begin{eqnarray}}
\newcommand{\eqr}{\end{eqnarray}}

\def\BC{\bb C}
\def\_\BC{\bbi C}

\def\( {\left(}
   \def\) {\right)}
\def\[ {\left[}
\def\] {\right]}
\def\no2 {{\textstyle{n\over 2}}}

\newcommand{\be}{\beta}

\newcommand{\pa}{\partial}

\newcommand{\da}{\dagger}

\newcommand{\lb}{\label}

\begin{document}
\begin{titlepage}
\setcounter{page}{1}
\renewcommand{\thefootnote}{\fnsymbol{footnote}}

\begin{flushright}
\end{flushright}

\vspace{5mm}
\begin{center}

{\Large \bf Goos-H\"{a}nchen like Shifts for Graphene Barrier\\ in
Constant Magnetic Field}

\vspace{5mm}

 {\bf Ahmed Jellal\footnote{\sf ajellal@ictp.it --
a.jellal@ucd.ac.ma}}$^{a,b}$, {\bf Miloud Mekkaoui}$^{b}$,
 and {\bf Youness Zahidi}$^{b}$

\vspace{5mm}

{$^a$\em Saudi Center for Theoretical Physics, Dhahran, Saudi Arabia}


{$^{b}$\em Theoretical Physics Group,  
Faculty of Sciences, Choua\"ib Doukkali University},\\
{\em PO Box 20, 24000 El Jadida, Morocco}


\vspace{3cm}

\begin{abstract}

We consider a system of Dirac fermions in graphene submitted to a
constant perpendicular magnetic field and scattered by a barrier
potential. We show that our system can be used to establish a link with
quantum optics through the Goos-H\"{a}nchen shifts. This can be done
by evaluating the corresponding transmission probability and shift phase.
We obtain Goos-H\"{a}nchen like  shifts in terms of different physical parameters
such as
 energy, electrostatic potential strength and  magnetic
field. On the light of this relation, we discuss the obtained results
and
make comparison with
literature.

\end{abstract}
\end{center}
\vspace{3cm}


\end{titlepage}

\section{Introduction}

Graphene~\cite{Novo, Novo2}, a single sheet of carbon honeycomb,
has ignited intense research activities to elucidate its
electronic properties. The linear energy dispersion relation of the
charge carriers, near the Dirac point in the electronic band
structure, make charge transport in graphene substantially
different from that of conventional two-dimensional (2D)
electronic systems~\cite{Nov, Zha}. What makes graphene so attractive for research
is that the spectrum closely resembles the Dirac spectrum for
massless fermions~\cite{Sem, Hal}. The charge carriers are
described as massless, chiral relativistic fermions, governed by
the Dirac equation. Their relativistic behavior  lead to one of
its fascinating properties is the so-called Klein
tunneling~\cite{Klein}. These properties are what enable graphene
to break so many records in terms of strength, electricity and
heat conduction as well as many others.
Electronic confinement
has been demonstrated in graphene microstructures using standard
lithography methods~\cite{Berg}, which makes the fabrication of
resonant-tunneling structures based on graphene practicable.

On the other hand, the Goos-H\"anchen (GH) effect \cite{Goos} is a phenomenon that
originated in classical optics in which a light beam reflecting
off a surface is spatially shifted as if it had briefly penetrated
the surface before bouncing back. The interface has to separate
different dielectric materials (such as glass or water), and
absorption or transmission should be small enough to allow a
substantial reflected beam to form \cite{Foster}. The size of the
GH effect is proportional to the derivative of the refection phase
with respect to the angle of incidence. In addition to shifting
beam position, the GH effect can manifest itself in alterations of
differential cross sections \cite{Tran} of laser mode dynamics
\cite{Dutriaux} and of mode spectra \cite{Chowdhurry}.

During the few past years there is a progress in studying
transport properties for charge carriers in graphene
through single or double magnetic barrier in combination with gate
voltage induced electrostatic potential.
Among these transport
properties we cite 
the quantum version of the GH effect originating from the reflection
of particles from interfaces.
Beenakker {\it et al.}~\cite{Beenakker} have pioneered the quantum GH
effect at the p-n interface in graphene. Many works in various
graphene-based nanostructures, including single~\cite{Chen},
double barrier~\cite{song, nous}, and superlattices~\cite{chenzh},
showed that the Goos-H\"anchen like (GHL) shifts can be enhanced by the transmission
resonances and controlled by varying the electrostatic potential
and induced gap~\cite{Chen}. Similar to those in semiconductors,
the GH shifts in graphene can also be modulated by electric and
magnetic barriers~\cite{sharma}, which influences the electronic
transport in graphene-based electronics. Also we mention that
the gate-controlled density of both p and n carrier types in
graphene was used to demonstrate the electronic analogue of fibre-optic
guiding \cite{tony}.



Very recently,
the GHL shifts for Dirac fermions in
graphene scattered by double barrier structures have been studied in \cite{nous}. After obtaining
the solution for the energy spectrum, the boundary
conditions have been used to explicitly determine GHL shifts
and the associated transmission probability. These two
quantities have been analyzed at resonances by studying their {main} characteristics
as a function of  the energy and electrostatic potential
parameters. {To check the validity of the computations done in \cite{nous}, the previous results
have been recovered
for a single barrier~\cite{Chen} under appropriate limits
and considerations.}

Motivated by different developments on the subject and in particular \cite{Chen,song, nous}, we investigate the GHL shifts
for
a system made of graphene with gap in presence of  constant magnetic field,
 single barrier potential and  Zeeman effect.
By splitting our system into three regions, we determine the solutions of the energy spectrum
in terms of different physical quantities.
After matching the wave functions
at both interfaces of potential width, we calculate the transmission coefficient
as well as the GHL shifts. To give a better understanding of our results, we plot different
figures versus physical parameters entering in the game and underline
their behaviors.


The present paper
is organized as follows. In section 2, we formulate our problem to
include different part in the Hamiltonian system describing
particles scattered by graphene barrier in external magnetic
field. We obtain the solution of the energy spectrum corresponding
to each region in terms of different scattering parameters. In
section 3 the scattering problem for Dirac fermions will be
solved, by using the boundary condition, in order to calculate the
transmission coefficient and their phase. In section 4, we analyze
the GHL shift and the transmission and we discuss our results by
presenting different plot. Finally, we conclude our results
and present some discussions.

\section{Theoretical model}

We consider a system of massless Dirac fermions through a strip of
graphene  with the Fermi energy $E$ and the incidence angle
$\phi_{0}$ with respect to the $x$-direction of the potential barrier.
This system is split into 
three regions denoted by 1, 2, 3 and each region is characterized
by the corresponding  potential and interaction with external sources
according to 
Figure \ref{barrier}:

\begin{figure}[H]
\centering
\includegraphics[width=10cm, height=6cm]{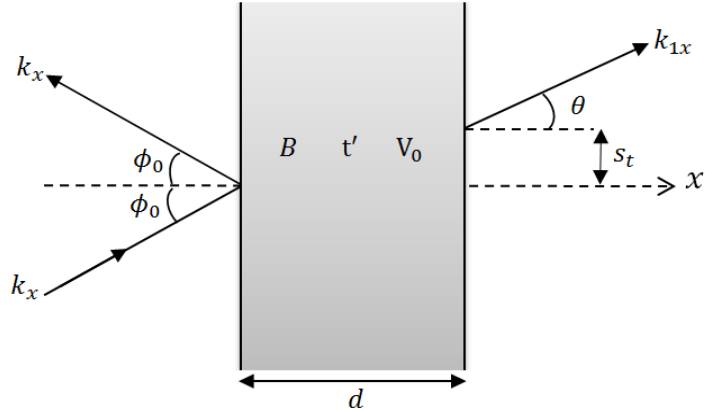}\\
\caption {\sf{ Schematic diagram for the monolayer graphene
barriers with different physical parameters.}}\label{barrier}
\end{figure}
\noindent The potential barrier has a rectangular shape
\begin{equation}
V(x)=
\left\{%
\begin{array}{ll}
     V_{0}, & \hbox{$0<x<d$} \\
    0, & \hbox{otherwise} \\
\end{array}%
\right.
\end{equation}
which is infinite
along the $y$-direction. 
In the barrier region of zero gap, both up-spin and down-spin
carriers obey the Dirac equation of motion described by the
Hamiltonian 
\beq
H_0=-i\hbar v_{F} \sigma \cdot
\nabla
\eeq
 with the linear dispersion relation $E=\hbar \epsilon
v_{F}$ where $v_{F}\approx 10^{6}ms^{-1}$ is the Fermi velocity
and $\sigma= (\sigma_{x}, \sigma_{y})$ represents the Pauli
matrices. In the central part of the sheet, we introduce an
external magnetic  field $B_{\circ}\hat{z}$, perpendicular to the graphene
sheet, along the $z$-direction defined by
\beq
B(x,y)=B_{0}\Theta(dx-x^{2})
\eeq
 where $\Theta$ is the Heaviside
step function. We also include the effect of Zeeman splitting, that
is characterized by an energy gap $2\Delta$. Choosing the Landau
gauge and imposing continuity of the vector potential at the
boundary to avoid nonphysical effects, the corresponding vector
potential $A(x,y)=(0,A_{y}(x))$ created by the magnetic field,
takes the  form
\begin{eqnarray}
\qquad A_{y}(x)=\frac{c}{el_{B}^{2}}\left\{%
\begin{array}{ll}
    0, & \hbox{$x<0$} \\
    x, & \hbox{$0\leq x\leq d$} \\
    d, & \hbox{$x>d$} \\
\end{array}%
\right.
\end{eqnarray}
where $l_{B}=\sqrt{c/eB_{0}}$ is the magnetic length, $c$ being
the speed of light and $e$ the electronic charge. After defining
the potential parameters relevant to all regions, we can write the
Hamiltonian for one-pseudospin component describing our system as
\begin{equation}\label{Htho}
H_{\tau}=v_{F} \vec\sigma\cdot \left(-i\hbar\vec\nabla+ \frac{e}{c}\vec{A}(x,
y)\right)+\Delta \sigma_{z}+V(x)+\frac{\tau}{2} g\mu_{B}B(x, y)
\end{equation}
where $\tau=1$ for up spin and $\tau=-1$ for down spin, the energy
gap is defined by
\beq
\Delta=t^{'}\Theta(dx-x^{2})
\eeq
the spin
$g$-factor (assumed to be 2) \cite{Zhang} of the carriers and
$\mu_{B}$ represents the Bohr magneton. Last term in \eqref{Htho}
is due to the Zeeman effect.

In order to solve the eigenvalue equation $H_\tau \psi=\epsilon \psi$ (in the unit system $\hbar$ = $c$ =$ e$ = 1), we can separate
variables and write the eigenspinors as
$\psi(x,y)=\varphi(x)\chi(y)$. We notice that the
momentum along the $y$-direction is conserved, which allows us to write
the eigenspinors as follows
\beq
\psi(x,y)=\varphi(x)e^{\tau i k_y y}.
\eeq
Then in region 1, we end up with
 the upper and lower components of
the eignespinor of the incident and reflection waves 
\begin{equation}
\psi_1(x,y)=\left(
\begin{array}{c}
1 \\
 \alpha\end{array}\right)e^{ik_{x}x}e^{\tau ik_{y}y} +  r_{\tau}\left(
\begin{array}{c}
1 \\
 -\alpha^{-1}\end{array}\right)e^{-ik_{x} x} e^{\tau ik_{y}
 y}
\end{equation}
where the complex parameter is
\begin{eqnarray*}
\alpha=s_{0}\frac{k_{x} +ik_{y}}{\sqrt{k_{x}^{2}
+k_{y}^{2}}}=s_{0} e^{\textbf{{i}}\phi}
\end{eqnarray*}
and $s_{0}=\mbox{\sf sgn}(\epsilon)$ refers to the conduction and
valence bands and $\phi$ is the angle that incident electrons make with the $x$-direction, see Figure \ref{barrier}.
The $x$- and $y$-components of the
electron wave vector, respectively, read as
\beq
 k_x=\epsilon \cos{\phi}, \qquad  k_y =\epsilon \sin{\phi}
 \eeq
which give the dispersion
relation 
\begin{equation}
\epsilon=s_{0} \sqrt{k^{2}_{x}+k_{y}^{2}}.
\end{equation}
In similar way to the above study, we can show that the wave function in the transmitted region 3 can
be expressed as
\begin{equation}
\psi_{3}(x,y)=t_{\tau}\left(
\begin{array}{c}
1 \\
 \beta\end{array}\right)e^{ik_{1x} x} e^{\tau ik_{y}
 y}
\end{equation}
where we have introduced the perpendicular and parallel
components of the wave vector as
\beq
k_{1x} = \epsilon
\cos{\theta} , \qquad k_y = \epsilon
\sin{\theta}-\frac{d}{l_{B}^2}
\eeq
and
  the parameter $\be$ is
\begin{eqnarray}
\beta=s_{0}\frac{k_{1x} +i
\left(k_{y}+\frac{d}{l^{2}_{B}}\right)}{\sqrt{k^{2}_{1x}
+\left(k_{y}+\frac{d}{l^{2}_{B}}\right)^{2}}}=s_{0}
e^{\textbf{\emph{i}}\theta}
\end{eqnarray}
with the phase
$\theta=\arctan{\left(\frac{k_y+{d}/{l_{B}^2}}{k_{1x}}\right)}$.
It is clearly seen that the dispersion relation takes the form
\begin{equation}
\epsilon=s_{0}
\sqrt{k^{2}_{1x}+\left(k_{y}+\frac{d}{l_{B}^2}\right)^{2}}.
\end{equation}
Now we can establish a connection between the two angles. This can fixed
by exploiting conservation of the momentum $p_y$ to end up with
\begin{equation}\label{}
    \sin{\theta}=\sin{\phi}+\frac{d}{\epsilon l_{B}^2 }.
\end{equation}
This will play a crucial role in the forthcoming analysis and especially
when we deal with the transmission and GH shifts.

To find all  solutions of the energy spectrum, we need to derive
from  \eqref{Htho} the Hamiltonian describing region 2 and
determine the corresponding eigenvalues and eigenspinors. Then, we
start from the eigenvalue equation
{$H_{\tau,2}\psi_{2}(x, y)= \epsilon \psi_{2}(x, y)$ }
with
\begin{equation}\label{H-2}
H_{\tau,2}=\left(%
\begin{array}{cc}
  v_{0}+\mu+\frac{\tau}{2l_{B}^{2}}g\mu_{0} & -i\left(\partial_{x}+k_{y}+\frac{x}{l_{B}^{2}}\right)\\
 i\left(-\partial_{x}+k_{y}
 +\frac{x}{l_{B}^{2}}\right)  &  v_{0}-\mu+\frac{\tau}{2l_{B}^{2}}g\mu_{0}\\
\end{array}%
\right), \qquad \psi_{2}(x, y)=(\varphi_1, \varphi_2)^{T}
\end{equation}
where we have set the parameters
 $v_{0}=V_{0}/\upsilon_{F}$, $\mu=t^{'}/\upsilon_{F}$,
$\mu_{0}=\mu_{B}/v_{F}$. Using all ingredients to obtain two
dependent first order differential equations
\begin{eqnarray}
  \left(v_{0}+\mu+\frac{\tau}{2l_{B}^{2}}g\mu_{0}\right)\varphi_{1}-i\frac{\sqrt{2}}{l_{B}}a\varphi_{2}=\epsilon\varphi_{1}\label{eq1}\\
  i\frac{\sqrt{2}}{l_{B}}a^{\dagger}\varphi_{1}+
  \left(v_{0}-\mu+\frac{\tau}{2l_{B}^{2}}g\mu_{0}\right)\varphi_{2}=\epsilon \varphi_{2} \label{eq2}
\end{eqnarray}
where the annihilation and creation operators are defined as
\begin{eqnarray}
a=\frac{l_{B}}{\sqrt{2}}\left(\partial_{x}+k_{y}+\frac{x}{l_{B}^{2}}\right),
\qquad
a^{\dagger}=\frac{l_{B}}{\sqrt{2}}\left(-\partial_{x}+k_{y}+\frac{x}{l_{B}^{2}}\right)
\end{eqnarray}
which satisfy the commutation relation $\left[a,a^\da\right]=\mathbb{I}$.
Now injecting \eqref{eq2} into \eqref{eq1} to find the second
differential equation for
 $\varphi_{1}$
\begin{equation}
\left[\left(\epsilon-v_{0}-\frac{\tau}{2l_{B}^{2}}g\mu_{0} \right)^{2}-\mu^{2}\right]\varphi_{1}=\frac{2}{l_{B}^{2}}a
a^{\dagger}\varphi_{1}.
\end{equation}
It is clear that $\varphi_{1}$ is an eigenstate of the number
operator $\widehat{N}=a^{\dagger}a$ and therefore we identify
$\varphi_{1}$ to the eigenstates of the harmonic oscillator
$ \mid n-1\rangle$, namely
\begin{equation}
 \varphi_{1} \sim \mid n-1\rangle
\end{equation}
which gives the energy spectrum
\begin{equation}
\epsilon-v_{0}-\frac{\tau}{2l_{B}^{2}}g\mu_{0}=s_1
k_{F}=s_1\frac{1}{l_{B}}\sqrt{(\mu l_{B})^{2}+2n}.
\end{equation}
The second spinor component can now be derived from
\eqref{eq2} as
\begin{equation}
\varphi_{2}=s_1 i\sqrt{\frac{k_{F}-s_1 \mu }{k_{F}+s_1 \mu }} \mid
n\rangle
\end{equation}
and therefore the corresponding eigenspinors, 
in terms of
 the parabolic cylinder functions
 \beq
D_{n}(x)=2^{-n/2}e^{\frac{-x^{2}}{4}}H_{n}\left(\frac{x}{\sqrt{2}}\right)
\eeq
can be written as
\begin{equation}
\psi_{2}(x, y)=\frac{1}{\sqrt{2}}\left(%
\begin{array}{c}
 \digamma_{1,\tau}
 D_{\left((k_F l_{B})^{2}-(\mu l_{B})^{2}
 \right)/2-1}
 \left[\pm \sqrt{2}\left(\frac{x}{l_{B}}+k_{y}l_{B}\right)\right] \\
  \pm i s_1 \digamma_{2,\tau}
  D_{\left((k_F l_{B})^{2}-(\mu l_{B})^{2}\right)/2}
  \left[\pm \sqrt{2}\left(\frac{x}{l_{B}}+k_{y}l_{B}\right)\right] \\
\end{array}%
\right)e^{\tau ik_{y}y}
\end{equation}
where the parameters $\digamma_{1}$ and $\digamma_{2}$ are
\begin{equation}
\digamma_{1,\tau}=\sqrt{\frac{ k_F+ s_1\mu }{ k_F}},\qquad
     \digamma_{2,\tau}=\sqrt{\frac{2}{ k_Fl_{B}^2( k_F+s_1\mu )}}
\end{equation}
and $\epsilon-V=s_1 k_F$, $s_1=\mbox{\sf sgn}\left(\epsilon-V\right)$ with
$V=v_{0}+\frac{\tau}{2l_{B}^{2}}g\mu_{0}$. The above $\pm$ signs
stand for waves traveling to right and left, respectively, along
the usual convention for the traveling waves $e^{\pm ik_{x}x}$ and
$e^{\pm ik_{1x} x}$. {The solution in region 2 is given by a linear
combination, $c^+$ with the upper sign added to $c^-$ with the
lower sign. As usual the arbitrary coefficients $c^\pm$ can be
determined using the boundary conditions, continuity of the
eigenspinors at each interface}. Next we will use the above
solutions to compute the transmission coefficient as well as the associated
phase shift and build a bridge between quantum optics and Dirac
fermions in graphene.

\section{Transmission and phase shift}

Before determining explicitly the transmission coefficient and the
phase shift, we notice that the total internal reflection will take
place only when $0<\phi<\frac{\pi}{2}$. This is because  the wave incident
from the right- and left-hand sides of the surface normal will
behave differently \cite{sank}. Then, we characterize our waves by
introducing a critical angle $\phi_c$ defined by
\begin{equation}\label{critic}
    \phi_c=\sin^{-1}{\left[1-\frac{d}{\epsilon l_B^2}\right]}.
\end{equation}
With this we can conclude that
when the incident angle is less than $\phi_c$, the
modes become oscillating guided modes. However, 
in the case when the
incident angle is more than $\phi_c$, we have decaying or
evanescent wave modes. In the forthcoming analysis, we will be
interested in studying the situation where $\phi<\phi_c$.

To deal with our task, we proceed by
using the continuity of the spinor wavefunctions at two interfaces $x=0$ and
$x=d$. These giving rise to a set of equations
\begin{eqnarray}
&& 1+r_{\tau}=
 c^{+}\digamma_{1,\tau}\eta_{1,\tau}^{+} +c^{-}\digamma_{1,\tau}\eta_{1,\tau}^{-}\\
&&
\alpha-\frac{1}{\alpha}r_{\tau}=c^{+}is_1\digamma_{2,\tau}\xi_{1}^{+}-c^{-}is_1\digamma_{2,\tau}\xi_{1}^{-}\\
&& c^{+}\digamma_{1,\tau}\eta_{2,\tau}^{+}+c^{-}\digamma_{1,\tau}\eta_{2,\tau}^{-}=t_{\tau}e^{ik_{1x}d}\\
 && c^{+}is_1\digamma_{2,\tau}\xi_{2,\tau}^{+} -c^{-}is_1\digamma_{2,\tau}\xi_{2,\tau}^{-}= \beta t_{\tau} e^{ik_{1x}d}.
\end{eqnarray}
After solving the system for the transmission amplitude $t_{\tau}$, which is
of interest to us, we can show that $t_{\tau}$ 
can be written in terms of the phase shift $\varphi$  as
\begin{equation}\lb{ttau}
t_{\tau}=\frac{e^{i\varphi_{\tau}}}{f_{\tau}}.
\end{equation}
Note that we have different reflection and transmission, which are depending on the nature of spin up or down.
The function $f_{\tau}$ 
can be obtained from the relation 
\begin{equation}\lb{ftau}
f_{\tau}e^{-i\varphi_{\tau}} =\frac{\left(A\cos\phi +B\sin
\phi\right)+i\left(B\cos\phi-A \sin\phi\right) }{
2F\cos\phi}e^{idk_{1x}}
\end{equation}
where the shorthand notation are defined by
\begin{eqnarray}
   A &=&s\digamma_{1,\tau}^{2} \left(\eta^{+}_{1,\tau}\eta^{-}_{2,\tau}-\eta^{-}_{1,\tau}\eta^{+}_{2,\tau}\right)\sin\theta+
      s\digamma_{2,\tau}^{2}\left(\xi^{-}_{1,\tau}\xi^{+}_{2,\tau}-\xi^{+}_{1,\tau}\xi^{-}_{2,\tau}\right)\sin\phi\\ \nonumber
        &&+s_1 \digamma_{1,\tau}\digamma_{2,\tau}\left[\left(\eta^{-}_{1,\tau}\xi^{+}_{2,\tau}+\eta^{+}_{1,\tau}\xi^{-}_{2,\tau}\right)
        +\left(\eta^{-}_{2,\tau}\xi^{+}_{1,\tau}+\eta^{+}_{2,\tau}\xi^{-}_{1,\tau}\right)\cos(\theta+\phi)\right]\\
  B &=&s\digamma_{1,\tau}^{2}\left(\eta^{-}_{1,\tau}\eta^{+}_{2,\tau}-\eta^{+}_{1,\tau}\eta^{-}_{2,\tau}\right)\cos\theta+
 s\digamma_{2,\tau}^{2}\left(\xi^{+}_{1,\tau}\xi^{-}_{2,\tau}-\xi^{-}_{1,\tau}\xi^{+}_{2,\tau}\right)\cos\phi\\ \nonumber
&&+s_1\digamma_{1,\tau}\digamma_{2,\tau}\left(\eta^{-}_{2,\tau}\xi^{+}_{1,\tau}+\eta^{+}_{2,\tau}\xi^{-}_{1,\tau}\right)\sin(\theta+\phi)
\end{eqnarray}
and we have  set the quantities
\begin{eqnarray}
    && \eta_{1,\tau}^{\pm}=D_{((k_f l_{B})^{2}-(\mu l_{B})^{2})/2-1} \left(\pm \sqrt{2}k_{y}l_{B}\right)\\
    &&\xi_{1,\tau}^{\pm}= D_{((k_f l_{B})^{2}-(\mu l_{B})^{2})/2} \left(\pm \sqrt{2}k_{y}l_{B}\right)\\
    && \eta_{2,\tau}^{\pm}=D_{((k_f l_{B})^{2}-(\mu l_{B})^{2} )/2-1} \left(\pm \sqrt{2} (\frac{d}{l_{B}}+k_{y}l_{B})\right)\\
    && \xi_{2,\tau}^{\pm}= D_{((k_f l_{B})^{2}-(\mu l_{B})^{2})/2} \left (\pm \sqrt{2}(\frac{d}{l_{B}}+k_{y}l_{B}) \right)\\
    &&F=s_1\digamma_{1}\digamma_{2} \left(\eta^{+}_{2,\tau}\xi^{-}_{2,\tau}+\eta^{-}_{2,\tau}\xi^{+}_{2,\tau}\right).
\end{eqnarray}
Actually from the above results, we can deduce the phase shift as
\begin{equation}
\varphi_{\tau}
=-\arctan\left[\frac{B\cos(\phi-k_{1x}d)-A\sin(\phi-k_{1x}d)}{A\cos(\phi-k_{1x}d)+B\sin(\phi-k_{1x}d)}\right]
\end{equation}
which can be used together with \eqref{ftau} to explicitly determine the transmission amplitude \eqref{ttau}. This statement will
be clarified in the forthcoming analysis.

Actually what we exactly  need are 
the reflection $R_{\tau}$ and transmission $T_{\tau}$ coefficients corresponding to our system.
These can be obtained by introducing
the electric current density $J$ corresponding to our system. Then from
the previous Hamiltonian, we show that $J$ takes the form 
\begin{equation}
J= \pm i\upsilon_{F}\psi^{\dagger}\sigma _{x}\psi
\end{equation}
which gives 
the incident, reflected and
transmitted current components
\begin{eqnarray}
        &&J_{\sf in}= \upsilon_{F}(\alpha+\alpha^{\ast})\\
        &&J_{{\sf re},\tau}= \upsilon_{F}r_{\tau}^{\ast}r_{\tau}(\alpha+\alpha^{\ast})\\
        &&J_{{\sf tr},\tau}= \upsilon_{F}t_{\tau}^{\ast}t_{\tau}(\beta+\beta^{\ast})
\end{eqnarray}
These can be used to define the reflection and transmission  coefficients as
\begin{equation}
     T_{\tau}=\frac{ |J_{{\sf tr},\tau}|}{| J_{\sf in}|},\qquad R_{\tau}=\frac{|J_{{\sf re},\tau}|}{ |J_{\sf in}|}
\end{equation}
where  $T^{\uparrow}$ and $T^{\downarrow}$ have similar
expression, the same applies to $R^{\uparrow}$ and
$R^{\downarrow}$, but the only difference being that $\tau=1$ instead of
$\tau=-1$.
After replacing, we obtain 
\begin{equation}
  T_{\tau}= \frac{k_{1x}}{k_x}|t_{\tau}|^{2}, \qquad
  R_{\tau}= | r_{\tau}|^{2}.
\end{equation}
 Combining all to
express the transmission probability as
\begin{equation}
    T_\tau=\frac{4k_{1x} F^2\cos^2\phi}{k_x\left(A^2+B^2\right)}.
\end{equation}
Obviously, we can check that 
the probability  condition $ T_\tau + R_\tau = 1$ is well satisfied. To under their behavior, let us
consider
Figure \ref{R+T}:
\begin{figure}[H]
\centering
\includegraphics[width=4in]{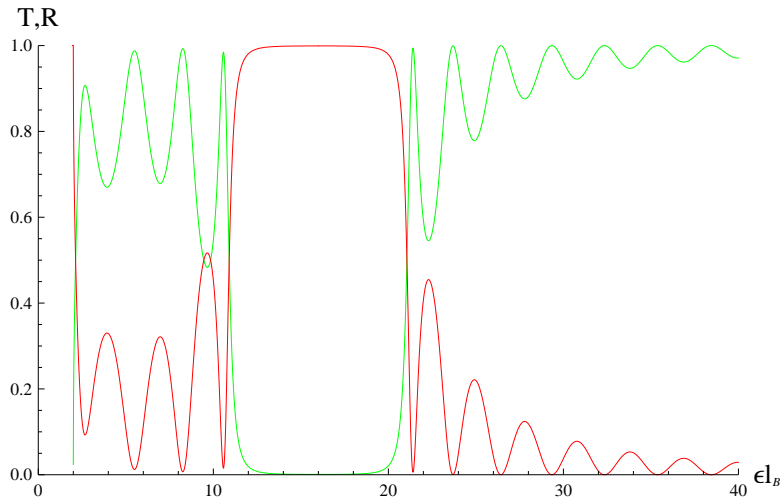}\\
\caption
 {\sf Transmission and reflection coefficients versus the energy $\epsilon l_B$ for a single barrier
 with the parameters:  $\mu_{0}=6l_{B}$, $d=1 l_B$, $v_{0}=\frac{10}{l_{B}}$,
 $\mu=\frac{4}{l_{B}}$ and $k_y=\frac{1}{l_{B}}$.}\label{R+T}
\end{figure}
\noindent
Figure \eqref{R+T} presents 
 the transmission (green line) and the reflection (red line) coefficients as function of the energy
 for $\mu_{0}=6l_{B}$, $d=1 l_B$, $v_{0}=\frac{10}{l_{B}}$,
 $\mu=\frac{4}{l_{B}}$ and $k_y=\frac{1}{l_{B}}$. Clearly, it shows that there are  different behaviors depending
 on the energy intervals. Indeed, in the first interval {$\left(\epsilon l_B<k_y l_B +\frac{d}{l_B}\right)$} we have no transmission because there
 is a forbidden zone \cite{jellal}. However, for the second interval $\left(k_y l_B +\frac{d}{l_B} <\epsilon l_B < v_0 l_B +\frac{1}{2 l_B}g \mu_0 -\mu \right)$,
 there are oscillation resonances due to the Klein regime that is the situation in which only oscillatory solutions exist
  throughout and where the so called Klein paradox reigns {\cite{Klein}}. We have no transmission (like a windows) when
  $v_0 l_B +\frac{1}{2 l_B}g \mu_0 -\mu<\epsilon l_B < v_0 l_B +\frac{1}{2 l_B}g \mu_0 +\mu$. Finally in the interval when
   $\epsilon l_B > v_0 l_B +\frac{1}{2 l_B}g \mu_0 +\mu$ contains the usual high energy barrier oscillations and
asymptotically goes to unity at high energy.

\section{Goos-H\"{a}nchen like shifts}

We begin our study of the Goos-H\"{a}nchen like (GHL) shifts by
considering an incident, reflected and transmitted beams, at
energy {$\epsilon$}, around some transverse wave
vector $k_y = k_{y_0}$ corresponding to the central incidence
angle $\phi_{0}$ (see Figure \ref{barrier}), denoted by the subscript $0$. These
beams can be expressed in the integral forms as
\begin{eqnarray}
 \psi_{\sf in}(x,y) &=&\int_{\infty}^{\infty}dk_{y}A(k_y)e^{i(k_{x}(k_y)x+\tau k_{y}y})\left(%
\begin{array}{c}
  1 \\
  s e^{i \phi(k_y)} \\
\end{array}%
\right)
\\
 \psi_{\sf re}(x,y) &=& \int_{\infty}^{\infty}dk_{y}r_\tau A(k_y)e^{i(-k_{x}(k_y)x+\tau k_{y}y})\left(%
\begin{array}{c}
  1 \\
  -s e^{-i \phi(k_y)} \\
\end{array}%
\right)
\\
 \psi_{\sf tr}(x,y) &=&\int_{\infty}^{\infty}dk_{y}t_{\tau}A(k_{y})e^{i(k_{1x}(k_y)+\tau k_{y}y})\left(%
\begin{array}{c}
  1 \\
  s_1 e^{i \theta(k_y)} \\
\end{array}%
\right).
\end{eqnarray}
The envelope function ensures the wavepacket is of finite size
along the $y$-direction and is sharply peaked at $k_y = k_{y_0}$.
Therefore, we have the intervals $k_{y_0}\in [0,\epsilon]$ and the angle of incidence
$\phi(k_y) \in\left[0,\frac{\pi}{2}\right]$. This fact is
represented by writing the $x$-component of wavevector $k_x$,
$k_{1x}$ as well as $\phi$ and $\theta$ all as function of $k_y$.
$A(k_y)$is the angular-spectrum distribution around the central
wave vector $k_{y_0}$, which can be assumed of the Gaussian shape
$w_ye^{-w_{y}^2(k_y-k_{y_0})^2}$ with $w_y$ being the half beam
width at waist \cite{Beenakker}. The reflection $r(k_y)$ and
transmission $t(k_y)$ coefficients will be calculated through the
use of boundary conditions. According to the stationary phase
approximation \cite{Chen}, the GHL shifts of the transmitted beam
through the graphene barrier can be defined as
\begin{equation}\label{st}
 s_{t}^{\tau}=-\frac{\partial\varphi_{\tau}}{\pa k_{y_0}}.
\end{equation}
We notice that the definition \eqref{st} can be applicable to any
finite-width beam, not necessarily a Gaussian-shaped beam. According to
\eqref{st}, we need first to
to determine the gradient of the phase $\varphi_{\tau}$ of the
transmission coefficient. 
Because the expression of $s_{t}^{\downarrow}$ is similar to that of spin up,
the only difference being that $\tau=1$
instead of $\tau=-1$, then in the forthcoming analysis we focus
only on 
$s_{t}^{\uparrow}$, which will simply be written as $s_t$.

\begin{figure}[H]
\centering
\includegraphics[width=8cm, height=6cm]{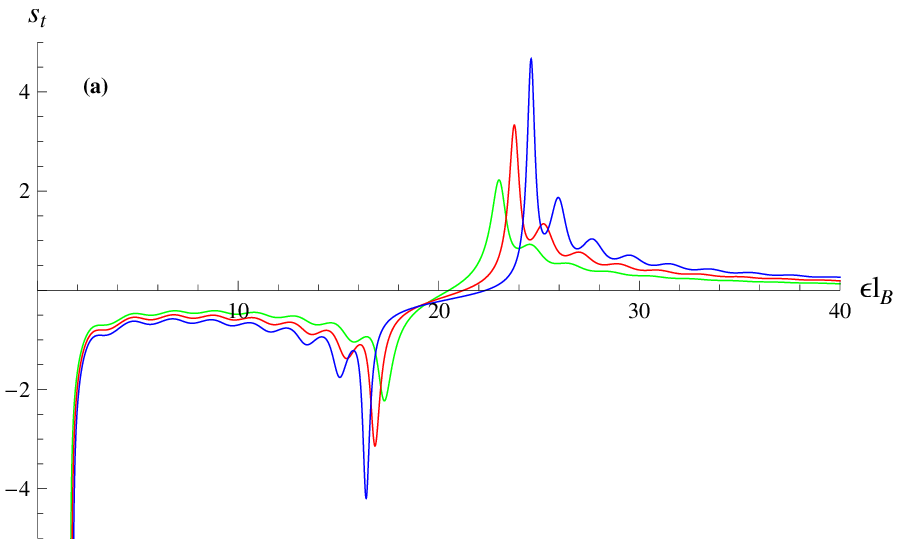}\ \ \ \ \ \ \
\includegraphics[width=8cm, height=6cm]{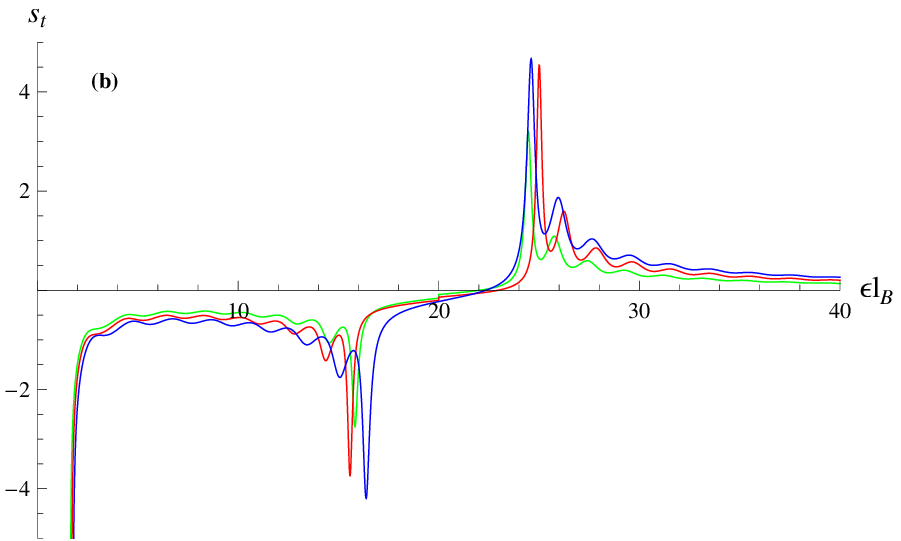}\\
 \includegraphics[width=8cm, height=6cm]{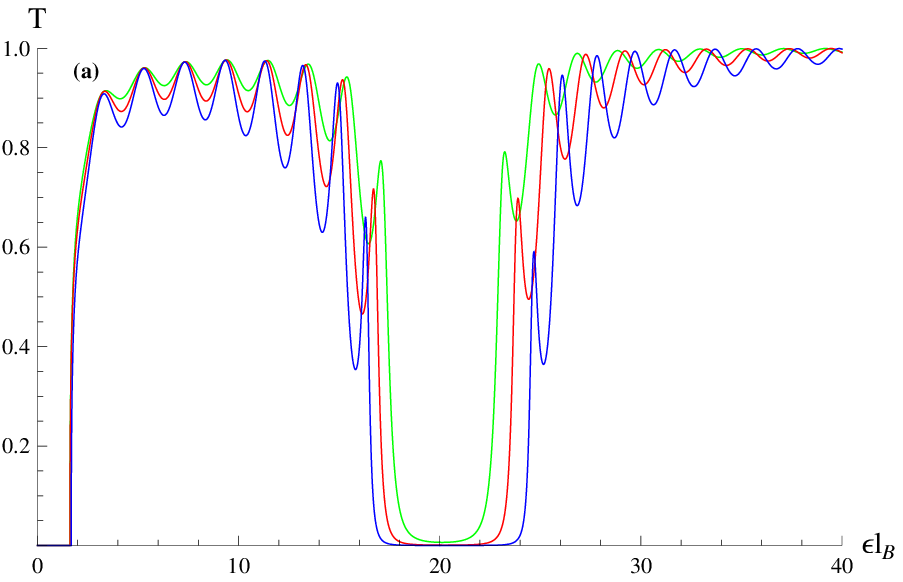}\ \ \ \ \ \ \
 \includegraphics[width=8cm, height=6cm]{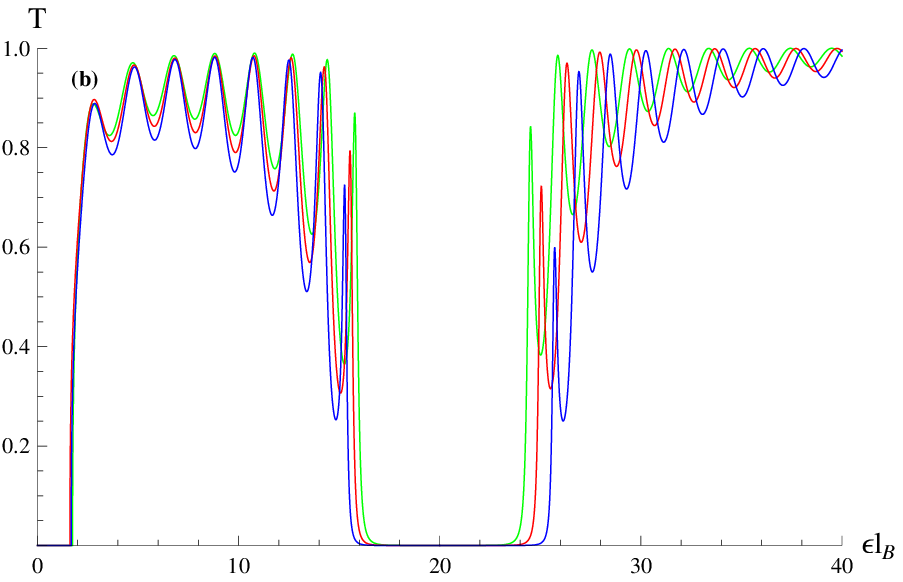}
 \caption
 {\sf {The GHL Shift and the transmission
  for graphene barrier in external magnetic field, as function of the energy(a)/(b), for $v_0=\frac{14}{l_B}$, $\mu_0=6l_B$, $d=1.5l_B$, $\tau=1$,
  $(\mu=\frac{0}{l_B}$, $\phi=4^\circ)$/$(\mu=\frac{3}{l_B}$, $\phi=4^\circ)$ (Green line),
   $(\mu=\frac{0}{l_B}$, $\phi=6^\circ)$/$(\mu=\frac{3}{l_B}$, $\phi=6^\circ)$ (Red line),
    $(\mu=\frac{0}{l_B}$, $\phi=8^\circ)$/$(\mu=\frac{3}{l_B}$, $\phi=8^\circ)$ (Blue line).}}\label{st+T-E}
\end{figure}

To allow for a
suitable interpretation of the GHL shift, we compute numerically
$s_t$ under various conditions. First we plot the GHL shifts as a
function of energy for different parameters of our system
considered in Figure \ref{barrier} ($v_0=14/l_B$, $\mu_0=6l_B$,
$d=1.5l_B$ and $\tau=1$) and three different values of the
incidence angle $\phi_1=4^\circ$, $6^\circ$, $8^\circ$ with
zero-gap $\mu=0$ (Figure \ref{st+T-E}.a) and finite gap
$\mu=3/l_B$ (Figure \ref{st+T-E}.b). Obviously, it is clearly seen
from Figure \ref{st+T-E} that the GHL shifts, for graphene
barrier in external magnetic field, can be negative as well as
positive and change the sign near the point $\epsilon
l_B=v_0l_B+\frac{1}{2l_B}g\mu_0$, whereas after certain threshold
energy value, which is compatible with a maximum of transmission,
the GHL shifts become constant. Another most impressive behavior of
the GHL shifts is that the position of the maximum or absolute value
of $s_t$ can be enhanced by the transmission resonances and
related to the transmission gap, as observed in single and double
barrier \cite{Chen, nous}. In addition to these properties, we
notice that the absolute values of the GHL shifts are strongly
dependent on the incidence angle, example of their independence is
plotted with zero and finite gap. From Figure \ref{st+T-E}, we can
show that the introduced gap influence the shifts. Then, it is also
interesting to discuss the influence of the induced gap.

Note that graphene is a zero gap semiconductor, despite these
great properties suitable, its one of the biggest hurdles for
graphene to be useful as an electronic material \cite{zhou},
however, to generate the energy gap is crucial for its application
in making devices. There are two ways to generate the energy gap
in monolayer graphene, one requires breaking of the translational
symmetry \cite{manes}, the other is to break the equivalence
between the A and B sublattice, which does not require any
translation symmetry breaking \cite{Trau, brey, son, nilsson}. These
fascinating properties of graphene suggest to underline the behavior our
findings with respect to the induced gap.

\begin{figure}[H]
\centering
\includegraphics[width=10cm, height=6cm]{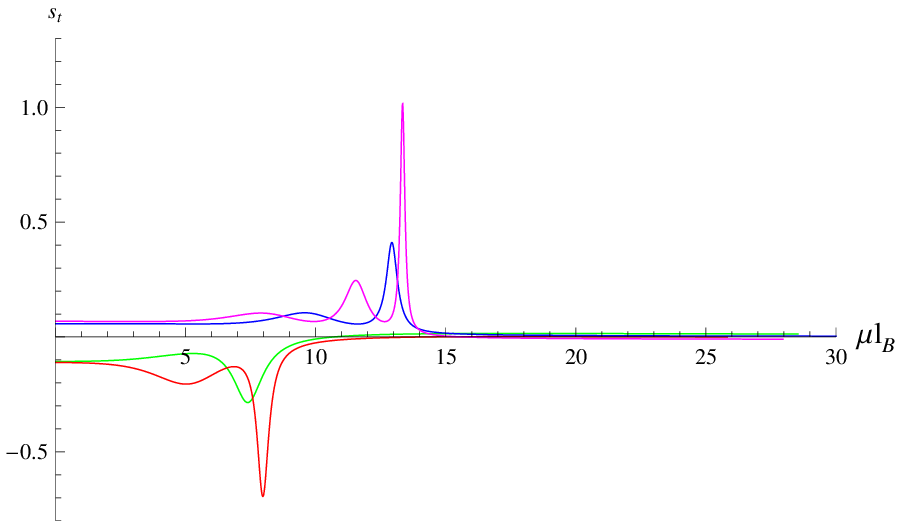}\\ \ \ \ \ \
 \includegraphics[width=10cm, height=6cm]{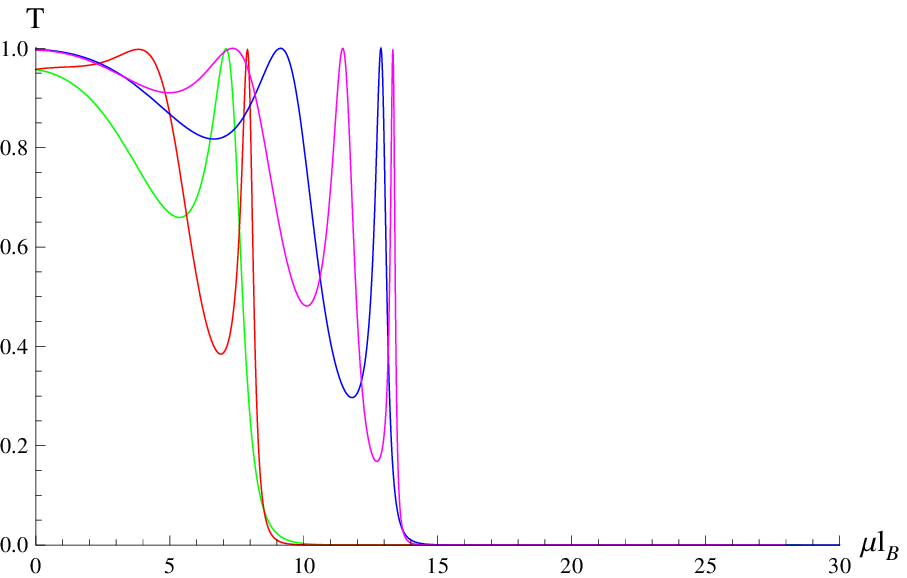} \caption
 {\sf {The influence of the induced gap on the GHL shifts and the transmission
  for graphene barrier in external magnetic field, for $v_0=\frac{15}{l_B}$,
  $\mu_0=6l_B$, $k_y=1$, $\tau=1$, $\epsilon=\frac{12}{l_B}$, $d=0.6 l_B$ (Green line), $\epsilon=\frac{12}{l_B}$, $d=0.8 l_B$ (Red line),
  $\epsilon=\frac{35}{l_B}$, $d=0.6 l_B$ (Blue line), $\epsilon=\frac{35}{l_B}$, $d=0.8 l_B$ (Purple
  line).}}\label{st+T-D}
\end{figure}

To answer above requirement, in Figure \ref{st+T-D} we show the influence of the induced gap on
the GHL shifts and the transmission as well. This has been performed by
fixing the parameters $v_0=\frac{15}{l_B}$, $\mu_0=6l_B$ and
making different choices of the energy and the barrier's width.
The GHL shifts become mostly constant up to some value then show
sharp picks. It is found that the GHL shifts can be enhanced by a
certain gap opening. In fact, we note that by increasing the gap,
the gap of transmission become boarder  that changing the
transmission resonances, where the modulation of the GHL shifts.
The dependence of the $s_t$ on the barrier's width is shown in
Figure \ref{st+T-D}. In addition, we  mention that in
the two configurations $\epsilon=\frac{12}{l_B}$, $d=0.6 l_B$
(Green line) and $\epsilon=\frac{12}{l_B}$, $d=0.8 l_B$ (Red line)
where the condition $\epsilon l_B<v_0 l_B+\frac{1}{2l_B}g\mu_0$ is
satisfied, the shifts is negative. However in the other two
configurations $\epsilon=\frac{35}{l_B}$, $d=0.6 l_B$ (Blue line)
and $\epsilon=\frac{35}{l_B}$, $d=0.8 l_B$ (Purple line) where
$\epsilon l_B>v_0 l_B+\frac{1}{2l_B}g\mu_0$, the shifts is
positive.

\begin{figure}[H]
\centering
\includegraphics[width=10cm, height=6cm]{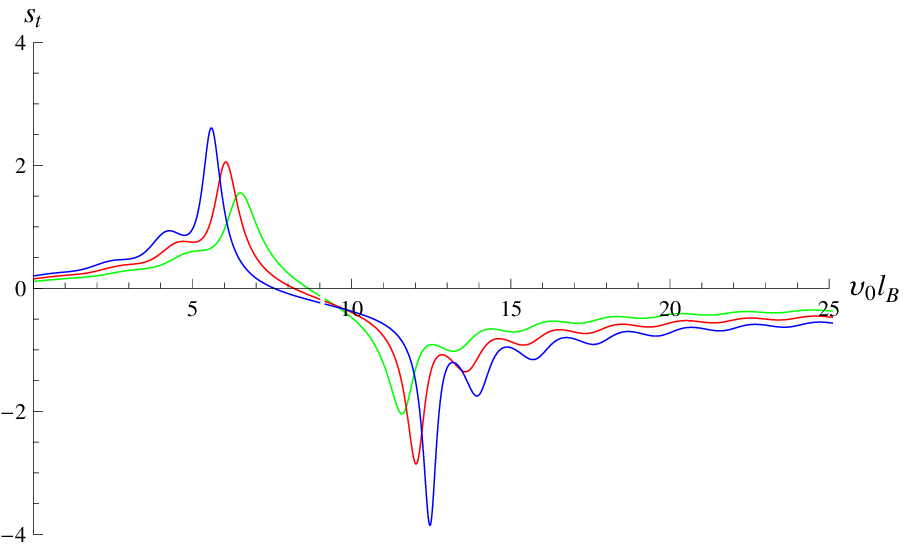}\\ \
 \includegraphics[width=10cm, height=6cm]{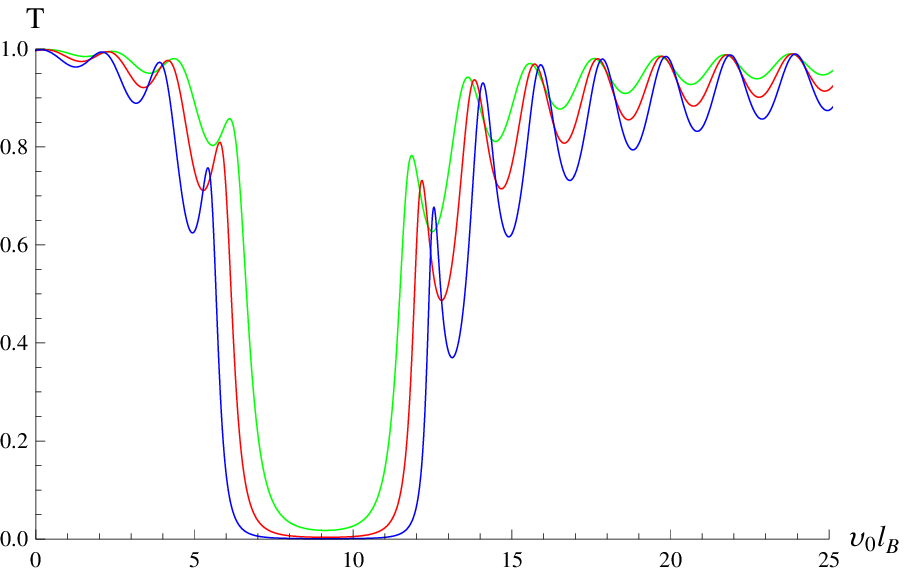}
 \caption {\sf {The GHL Shift and the transmission
  for graphene barrier in external magnetic field as function of the height of the potential barrier, for $\epsilon=\frac{15}{l_B}$, $\mu_0=6l_B$, $d=1.5l_B$,
  $\tau=1$,$\mu=\frac{0}{l_B}$, $\phi=4^\circ$ (Green line), $\phi=6^\circ$ (Red line), $\phi=8^\circ$ (Blue line).}}\label{st+T-v0}
\end{figure}

Now let us investigate how the GHL shift behave as function of the
barrier potential height that is represented numerically in
Figure \ref{st+T-v0}. We have fixed the parameters
$\epsilon=\frac{15}{l_B}$, $\mu_0=6l_B$,
$d=1.5l_B$,$\mu=\frac{0}{l_B}$ and made different choices for
the incidence angle $\phi=4^\circ, 6^\circ, 8^\circ$. It is
clearly shown that the GHL shifts change their sign near the point
$v_0l_B=\epsilon l_B-\frac{1}{2l_B}g\mu_0$ and behave differently
as compared to Figure \ref{st+T-E}. We notice that when the
condition $v_0l_B<\epsilon l_B-\frac{1}{2l_B}g\mu_0$ is fulfilled, the GHL
shifts are positive, while  negative when the height of the
barrier satisfies the condition $v_0l_B>\epsilon
l_B-\frac{1}{2l_B}g\mu_0$. It is clearly seen from
Figure \ref{st+T-v0} that the GHL shifts are strongly dependent on
the barrier  height $V_0$, which experimentally can be implemented  
by applying a local top gate voltage $V_0$ to graphene \cite{huard}.
This tells us  that the GHL shifts can be controlled by changing
$V_0$.\\

\begin{figure}[H]
\centering
\includegraphics[width=10cm, height=6cm]{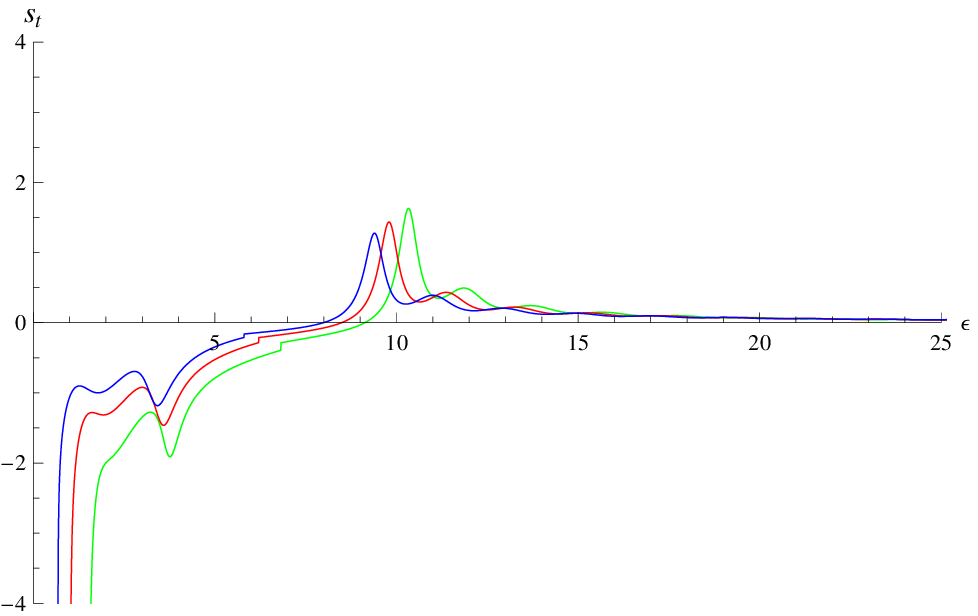}\\ \
 \includegraphics[width=10cm, height=6cm]{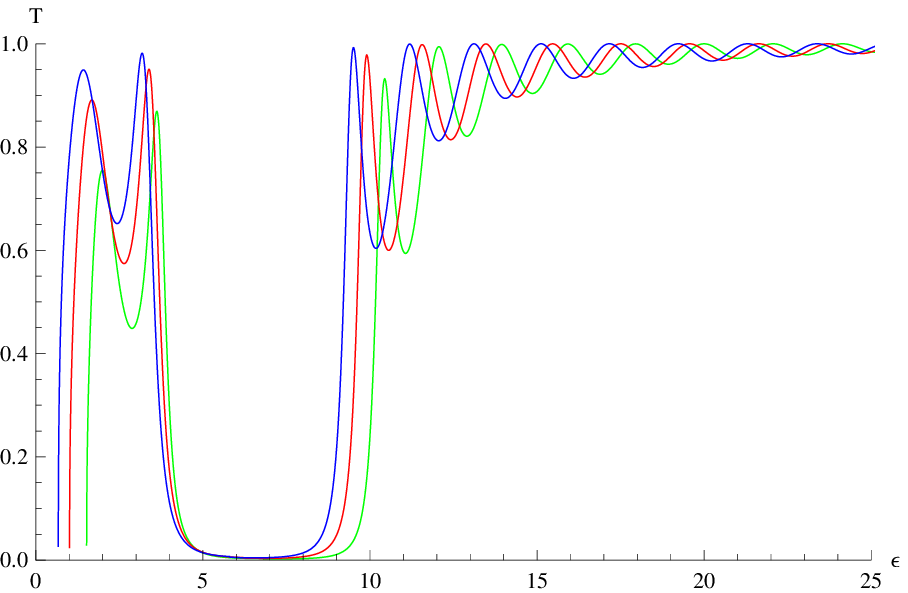}
 \caption {\sf {The GHL Shift and the transmission
  for graphene barrier in external magnetic field versus the energy, for $V_0=5meV$, $d=1.5
  nm$, { $ t'=2meV$}, $\phi=6^\circ$ and
  $B=0.9T$ (Green line), $B=0.6T$ (Red line), $B=0.4T$ (Blue line).}}\label{st+T-E-lb}
\end{figure}

In order to investigate what will happen when we change the value
of the magnetic field, we plot the GHL shifts as function of the
energy with different value of the magnetic field. The numerical
results are shown in Figure \ref{st+T-E-lb} with the parameters
$V_0=6meV$, $d=1.5nm$, {$t'=2meV$}, $\phi=6^\circ$ and
different configurations of $B$ ($0.9T$, $0.6T$, $0.4T$). We
conclude that when we increase $B$ the transmission gap become
larger, the GHL shifts still change the sign, but  the point of
change of the sign moves to the left and the absolute value of the
maximum of the shifts increase.

\section{Conclusion}

We have considered Dirac fermions in graphene with gap
and submitted to a constant magnetic field as well as
a barrier potential. By including Zeeman effect,
the solutions of the energy spectrum were obtained to be dependent on
various physical parameters. By matching the eigenspinors at interfaces
we have calculated
the transmission amplitude in terms of the phase shift $\varphi_\tau$.
These is used together with current density of our system to
determine transmission and reflections probabilities.
the transmission and reflections probabilities. Their behaviors
were characterized by oscillating resonances for some interval of energy
and forbidden zone where the reflection is total.

Subsequently, we have computed the Goos-H\"{a}nchen like shifts for
graphene barrier in external magnetic field $B$. This has been done
by considering an incident, reflected and
transmitted beams, at energy $\epsilon$, around some transverse wave vector $k_y = k_{y_0}$ corresponding to the
central incidence angle $\phi_{0}$. We have given different plots of the shifts and the corresponding
transmission coefficient versus different physical parameters.
In fact, we have observed that there is a strong dependence of the GHL shifts on the
incidence angle $\phi_0$ (see Figure \ref{barrier}) and change as long as $\phi_0$ changes.
Also, the magnetic field affected the GHL shifts 
shifts because when  $B$ is increased,
transmission gap
became larger, the GHL shifts still change the sign, but the point of change of the sign moved to the
left and the absolute value of the maximum of the shifts increased as well.
It was also observed
that the transmission gap increased with the angle of incidence as
long as it was less than the critical angle.

By comparing our
results with those found for single and double barriers without $B$ field \cite{nous,
Chen,song}, we have conclude that the common feature is that the GHL shifts can be positive
as well as negative and can be enhanced by transmission
resonances. Also  the negative or positive shifts can be
modulated by the height and width of potential barrier and the
induced gap. These investigations
allow to have potential application
in various graphene based electronic devices.

\section*{Acknowledgment}

The generous support provided by the Saudi Center for Theoretical
Physics (SCTP) is highly appreciated by all authors.


\begin{thebibliography}{99}

\bibitem{Novo}    K. S. Novoselov, A. K. Geim, S. V. Morozov, D. Jiang, Y. Zhang, S. V. Dubonos, I. V. Grigorieva and A. A. Firsov, {Science} {306}, 666 (2004).

\bibitem{Novo2}   K. S. Novoselov, D. Jiang, T. Booth, V. V . Khotkevich, S. M. Morozov and A. K. Geim,  Proc. Natl. Acad. Sci. 102, 10451 (2005).

\bibitem{Nov}     K. S. Novoselov, A. K. Geim, S. V. Morozov, D. Jiang, M. I. Katsnelson, I. V. Grigorieva, S. V. Dubonos and A. A. Firsov, Nature 438, 197 (2005).

\bibitem{Zha}     Y. Zhang, Y. W. Tan, H. L. Stormer and P. Kim, Nature 438, 201 (2005).

\bibitem{Sem}     G. W. Semenoff, Phys. Rev. Lett. 53, 2449 (1984).

\bibitem{Hal}     F. D. M. Haldane, Phys. Rev. Lett. 61, 2015 (1988).

\bibitem{Klein}   O. Klein,  Z. Phys. 53, 157 (1929).



\bibitem{Berg}    C. Berger, Z. Song, X. Li, X. Wu, N. Brown, C. Naud, D. Mayou, T. Li, J. Hass, A. N. Marchenkov, E. H. Conrad, P. N. First and W. A. de Heer, Science 312, 1191 (2006).

\bibitem{Goos}    F. Goos and H. H\"anchen,  Ann. Phys. 436, 333 (1947).

\bibitem{Foster}  D. H. Foster, J. U. N\"ockel and A. K. Cook, Opt. Lett. 32, 1764 (2007).

\bibitem{Tran}    N. H. Tran, L. Dutriaux, P. Balcou, A. Le Floch and F. Bretenaker,  Opt. Lett. 20, 1233 (1995).

\bibitem{Dutriaux} L. Dutriaux, A. Le Floch and F. Bretenaker, J. Opt. Soc. Am. B 12, 2283 (1992).

\bibitem{Chowdhurry} D. Q. Chowdhurry, D. H. Leach and R. K. Chang,  J. Opt. Soc. Am. A 11, 1110 (1994).


\bibitem{Beenakker}  C. W. J. Beenakker, R. A. Sepkhanov, A. R. Akhmerov  and J. Tworzydlo,   Phys. Rev. Lett. 102, 146804 (2009).

\bibitem{Chen}    X. Chen, J.-W. Tao, and Y. Ban, Eur. Phys. J. B 79, 203  (2011).

\bibitem{song}    Y. Song, H-C. Wu  and Y. Guo,   Appl. Phys. Lett. 100, 253116 (2012).

\bibitem{nous}    A. Jellal, I. Redouani, Y. Zahidi and H. Bahlouli, Physica E 58, 30 (2014). 

\bibitem{chenzh}  X. Chen, P-L. Zhao, X-J. Lu  and L-G. Wang, Eur. Phys. J. B 86, 223 (2013). 

\bibitem{sharma}  M. Sharma  and S. J. Ghosh,  Phys.: Condens. Matter 23, 055501 (2011).

\bibitem{tony} J. R. Williams, Tony Low, M. S. Lundstrom and C. M. Marcus, Nature Nanotechnology 6, 222 (2011).

\bibitem{Zhang}   Y. Zhang, Z. Jiang, J. P. Small, M. S. Purewal, Y.-W. Tan, M. Fazlollahi, J. D. Chudow, J. A. Jaszczak, H. L. Stormer
 and P. Kim, Phys. Rev. Lett. 96, 136806 (2006).

 \bibitem{sank}    S. Ghosh and M. Sharma, J. Phys.: Condens. Matter 21, 292204 (2009).

 \bibitem{jellal} A. Jellal, E. B. Choubabib, H. Bahlouli and A. Aljaafari,
J. Low Temp. Phys. 168, 51 (2012).


 \bibitem{zhou}    S. Y. Zhou, G. H. Gweon, A. V. Federov, P. N. First, W. A. de Heer, D. H. Lee, F. Guinea, A. H. Castro Neto
 and A. Lanzara, Nat. Mater. 6, 770 (2007).

\bibitem{manes}   J. L. Manes, F. Guinea  and  A. H. Vozmediano,  Phys. Rev. B 75, 155424 (2007).

\bibitem{Trau}    B. Trauzettel, D. V. Bulaev, D. Loss  and G. Burkard,  Nature Phys. 3, 192 (2007).

\bibitem{brey}    L. Brey and H. A. Fertig,  Phys. Rev. B 73, 235411 (2006).

\bibitem{son}     Y. W. Son, M. L. Cohen,  and S. G. Louie, Phys. Rev. Lett. 97, 216803 (2006).

\bibitem{nilsson} J. Nilsson, A. H. Castro Neto, F. Guinea and  N. M. R. Peres, Phys. Rev. B 76, 165416 (2007).

\bibitem{huard}   B. Huard, J. A. Sulpizio, N. Stander, K. Todd, B. Yang and D. Goldhaber-Gordon, Phys. Rev. Lett. 98, 236803 (2007).









\end{thebibliography}
\end{document}